\documentclass[12pt]{article}
\usepackage{latexsym}
\usepackage{ifthen}
\usepackage{epsf}

\def\hybrid{\topmargin -20pt    \oddsidemargin 0pt
        \headheight 0pt \headsep 0pt
        \textwidth 6.5in        % US paper
        \textheight 9in         % US paper
        \marginparwidth .875in
        \parskip 5pt plus 1pt   \jot = 1.5ex}

\hybrid
\makeatletter
\@addtoreset{equation}{section}
\makeatother

\newcommand{\color}[6]{}

\newcommand{\cB}{{\cal B}}

\newcommand{\cM}{{\cal M}}
\newcommand{\cN}{{\cal N}}

\newcommand{\cQ}{{\cal Q}}

\newcommand{\hf}{\frac12}

\newcommand{\bea}{\begin{eqnarray}}
\newcommand{\eea}{\end{eqnarray}}
\newcommand{\be}{\begin{equation}}
\newcommand{\ee}{\end{equation}}
\newcommand{\bt}{\begin{tabular}}
\newcommand{\et}{\end{tabular}}
\newcommand{\ba}{\begin{array}}
\newcommand{\ea}{\end{array}}

\newcommand{\vev}[1]{\langle #1 \rangle}

\newcommand{\Tr}{\mathop{\rm Tr}}
\newcommand{\tr}{\mathop{\rm tr}}

\newcommand{\cof}{\mathop{\rm cof}}

\def\IB{\relax{\rm I\kern-.18em B}}
\def\ID{\relax{\rm I\kern-.18em D}}
\def\IE{\relax{\rm I\kern-.18em E}}
\def\IF{\relax{\rm I\kern-.18em F}}
\def\IH{\relax{\rm I\kern-.18em H}}
\def\II{\relax{\rm I\kern-.18em I}}
\def\IK{\relax{\rm I\kern-.18em K}}
\def\IL{\relax{\rm I\kern-.18em L}}
\def\IM{\relax{\rm I\kern-.18em M}}
\def\IN{\relax{\rm I\kern-.18em N}}
\def\IP{\relax{\rm I\kern-.18em P}}
\def\IR{\relax{\rm I\kern-.18em R}}
\def\IT{\relax{\rm I\kern-.42em T}}
\def\IZ{\relax{\hbox{\raisebox{.38ex}
    {\scriptsize\bfseries\slshape /}\kern-.40em\_\kern-.28em\rm Z}}}
\def\Iz{\relax{\hbox{\raisebox{.38ex}
    {\tiny\bfseries\slshape /}\kern-.25em\raisebox{.65ex}
    {\tiny\bfseries\slshape /}\kern-.43em\_\kern-.26em\rm Z}}}
\def\inbar{\vrule height1.5ex width.8pt depth-0.2pt}
\def\inbarhi{\vrule height1.55ex width.5pt depth-.85ex}
\def\inbarlo{\vrule height.8ex width.5pt depth0ex}
\def\IC{\relax{\rm C\kern-.48em \inbar\kern.48em}}
\def\IO{\relax{\rm O\kern-.56em \inbar\kern.56em}}
\def\IQ{\relax{\rm Q\kern-.56em \inbar\kern.56em}}
\def\IS{\relax{\rm S\kern-.37em \inbarhi\kern.08em\inbarlo\kern.29em}}

\def \one{\hbox{\bf 1}}

\def \half{{\textstyle\hf}}

\def \soll={\stackrel{!}{=}}

\def \Wtree{W_{\rm tree}}
\def \Weff{W_{\rm eff}}
\def \Wconf{W_{\rm conf}}

\def \first{$1^{\rm st}$}
\def \second{$2^{\rm nd}$}

\def \barfill{\leaders\hrule height 0.1 true pt\hfill}
\def \overbar#1{\vbox{\ialign{##\crcr\barfill\crcr\noalign{\kern 1pt
                                      \nointerlineskip}$\hfil{#1}\hfil$\crcr}}}
\def \scriptbar#1{{\vbox{\ialign{##\crcr\thinspace\barfill\thinspace\crcr
    \noalign{\kern 0.8pt\nointerlineskip}$\hfil{\scriptstyle #1}\hfil$\crcr}}}}

\newlength{\oldindent}
\newlength{\quadlength} 
\newlength{\abstand} \newlength{\breite}

\newcommand{\drawsquare}[2]{\hbox{%
\rule{#2pt}{#1pt}\hskip-#2pt%  left vertical
\rule{#1pt}{#2pt}\hskip-#1pt%  lower horizontal
\rule[#1pt]{#1pt}{#2pt}}\rule[#1pt]{#2pt}{#2pt}\hskip-#2pt%  upper horizontal
\rule{#2pt}{#1pt}}% right vertical

\newcommand{\Yfun}{\raisebox{-.5pt}{\drawsquare{6.5}{0.4}}}%  fund
\def \Yfunb{\overline{\Yfun}}

\def\cnodea{\put(0.8,2.5){\circle*{1.6}}}
\def\cnodeb{\put(7.4,2.5){\circle*{1.6}}}
\def\rnodea{\put(0.8,2.5){\circle{1.6}}}
\def\rnodeb{\put(7.4,2.5){\circle{1.6}}}
\def\pnodea{\put(0.8,2.5){\circle{1.6}}\put(0.8,2.5){\circle*{0.4}}}
\def\pnodeb{\put(7.4,2.5){\circle{1.6}}\put(7.4,2.5){\circle*{0.4}}}
\newlength{\firstlength} \newlength{\secondlength}
\def\qlink#1#2{\begin{picture}(9,4)
            \def\first{#1} \def\second{#2}
            \settowidth{\firstlength}{$k$} \settowidth{\secondlength}{$l$}
            \addtolength{\firstlength}{-0.5\firstlength}
            \addtolength{\secondlength}{-0.5\secondlength}
            \def\cx{c} \def\rx{r} \def\px{p}
            \ifx\first\cx \cnodea \else
               \ifx\first\rx \rnodea \else \pnodea \fi\fi
            \ifx\second\cx \put(1.6,2.5){\vector(1,0){5}}\cnodeb \else
               \put(1.6,2.5){\line(1,0){5}} \ifx\second\rx \rnodeb \else
               \pnodeb\fi\fi
            \put(1,-0.9){\hspace{-\firstlength}\scriptsize$k$}
            \put(7.6,-0.9){\hspace{-\secondlength}\scriptsize$l$}
            \end{picture}}
\def\qlinkx#1#2#3#4#5#6{\begin{picture}(9,4)
            \def\tail{#1} \def\head{#2}
            \def\first{#3} \def\second{#4}
            \settowidth{\firstlength}{$#5$} \settowidth{\secondlength}{$#6$}
            \addtolength{\firstlength}{-0.5\firstlength}
            \addtolength{\secondlength}{-0.5\secondlength}
            \def\cx{c} \def\rx{r} \def\px{p} \def\yes{1} \def\no{0}
            \ifx\first\cx \cnodea \else
               \ifx\first\rx \rnodea \else \pnodea \fi\fi
            \if\head\yes \put(1.6,2.5){\vector(1,0){5}} 
               \else\ifx\tail\no \put(1.6,2.5){\line(1,0){5}} \fi\fi
            \if\tail\yes \put(6.6,2.5){\vector(-1,0){5}} \fi
            \ifx\second\cx \cnodeb \else
               \ifx\second\rx \rnodeb \else \pnodeb\fi\fi
            \put(1,-0.9){\hspace{-\firstlength}\scriptsize$#5$}
            \put(7.6,-0.9){\hspace{-\secondlength}\scriptsize$#6$}
            \end{picture}}
\def\qtens#1#2{\def\first{#1}\def\cx{c}\def\rx{r}\def\px{p}%
               \settowidth{\firstlength}{$#2$}
               \addtolength{\firstlength}{-0.5\firstlength}
             \ifx\first\cx 
                   \begin{picture}(3,6)
                     \cnodea
                     \put(0.8,4.5){\circle{4}}
                     \put(1,-0.9){\hspace{-\firstlength}\scriptsize$#2$}
                   \end{picture}%
              \else% 
                   \begin{picture}(3,6)
                     \ifx\first\rx\rnodea\else\pnodea\fi
                     \qbezier(-0.1,3)(-1.2,3.72843)(-1.2,4.5)
                     \qbezier(-1.2,4.5)(-1.2,5.32843)(-0.61421,5.91421)
                     \qbezier(-0.61421,5.91421)(-0.02843,6.5)(0.8,6.5)
                     \qbezier(0.8,6.5)(1.62843,6.5)(2.21421,5.91421)
                     \qbezier(2.21421,5.91421)(2.8,5.32843)(2.8,4.5)
                     \qbezier(2.8,4.5)(2.8,3.6)(1.6,3)
                     \put(1,-0.9){\hspace{-\firstlength}\scriptsize$#2$}
                   \end{picture}%
              \fi}

\newlength{\totallength}
\newlength{\height}
\newlength{\firstheight}\newlength{\secondheight}
\newlength{\ruleheight}\newlength{\midruleheight}
\newlength{\lineheight}\newlength{\hcorrection}
\newcommand{\contraction}[3][0em]{\setlength{\hcorrection}{#1}
                \settowidth{\firstlength}{$#2$} 
                \settowidth{\secondlength}{$#3$}
                \settoheight{\firstheight}{$#2$}
                \settoheight{\secondheight}{$#3$}
                \setlength{\lineheight}{0.3pt}
                \setlength{\ruleheight}{1ex}
                \setlength{\midruleheight}{\ruleheight}
                \addtolength{\midruleheight}{-\lineheight}
                \addtolength{\firstlength}{-0.5\firstlength}
                \addtolength{\secondlength}{-0.5\secondlength}
                \setlength{\totallength}{\firstlength}
                \addtolength{\totallength}{\secondlength}
                \ifthenelse{\firstheight<\secondheight}
                           {\setlength{\height}{\secondheight}}
                           {\setlength{\height}{\firstheight}}
                \addtolength{\height}{0.25ex}
                \hspace*{\firstlength}\hspace*{\hcorrection}
                \rule[\height]{\lineheight}{\ruleheight}
                \addtolength{\height}{\midruleheight}
                \rule[\height]{\totallength}{\lineheight}
                \addtolength{\height}{-\midruleheight}
                \rule[\height]{\lineheight}{\ruleheight}
                \hspace{-\firstlength}\hspace{-\totallength}
                \hspace*{-\hcorrection}
                #2#3}

\renewcommand{\thefootnote}{\fnsymbol{footnote}}

\begin{document}

%----------------------------------------------------------------------%
%  title page
%----------------------------------------------------------------------%
\begin{titlepage}
\begin{center}

\rightline{}
\rightline{SLAC-PUB-10156}
\rightline{hep-th/0309044}

\vskip .6in
{\LARGE \bf On effective superpotentials and Kutasov duality}
\vskip .8in

{\bf Matthias Klein\footnote{E-mail: mklein@slac.stanford.edu},
     Sang-Jin Sin\footnote{E-mail: sjsin@hanyang.ac.kr}}
\vskip 0.8cm
{\em $^{*\,\dagger}$ Theory Group, SLAC, Stanford University, Stanford, CA 94309, USA.}
\vskip 0.2cm
{\em  
$^\dagger$ Department of Physics, Hanyang University, Seoul, 133-791, Korea.}
\end{center}

\vskip 1.5cm

\begin{center} {\bf ABSTRACT } \end{center}
We derive the effective superpotential for an $\cN=1$ $SU(N_c)$ 
gauge theory with one massless adjoint field and $N_f$ massless
fundamental flavors and cubic tree-level superpotential for the 
adjoint field. This is a generalization of the Affleck-Dine-Seiberg 
superpotential to gauge theories with one massless adjoint matter 
field.
Using Kutasov's generalization of Seiberg duality, we then find the 
effective superpotential for a related theory with massive fundamental 
flavors.

\vfill

September 2003
\end{titlepage}

\newpage

%----------------------------------------------------------------------%
%  Resetting of counters
%----------------------------------------------------------------------%
\setcounter{page}{1} \pagestyle{plain}
\renewcommand{\thefootnote}{\arabic{footnote}}
\setcounter{footnote}{0}

%----------------------------------------------------------------------%
%  Paper begins
%----------------------------------------------------------------------%

\section{Introduction}
Nonperturbative superpotentials are important because they lead to 
an understanding of interesting quantum effects in supersymmetric 
field theories and because of their phenomenological implications for 
supersymmetric models. Furthermore, they are an ideal arena to test 
our computational methods in supersymmetric field theories.  
We will use a combination of several nonperturbative methods
to determine the effective superpotential for a class of
supersymmetric gauge theories with matter in the adjoint and
fundamental representations.
In earlier days, Affleck, Dine and Seiberg \cite{ADS}
were able to determine the effective nonperturbative
superpotential for SQCD with $N_f<N_c$, this is an $\cN=1$ 
$SU(N_c)$ gauge theory with $N_f$ quark flavors, by performing an 
explicit instanton calculation.

During the last decade many powerful techniques have been 
developed to compute effective superpotentials for broad classes
of supersymmetric field theories. Seiberg showed \cite{Sei} that
in many cases the nonperturbative superpotential is completely
determined by the symmetries and some physical boundary conditions.
He also found \cite{Sei2} that the 't Hooft anomaly matching 
conditions \cite{tHooft} are an important constraint when 
determining the low-energy spectrum of gauge invariant operators
and used this to establish an electric-magnetic duality \cite{Sei3}
relating two different $\cN=1$ gauge theories in the far IR.
Some of these results were subsequently generalized to supersymmetric
gauge theories with different gauge groups and/or matter content.
Of particular interest is a model first discussed by Kutasov
\cite{Ku} containing a massless adjoint field and 
massless fundamental fields.
However, although many nonperturbative results for the Kutasov 
model have been obtained, including a dual description in terms 
of magnetic variables \cite{Ku,KS,KSS}, an effective superpotential 
generalizing the result of Affleck, Dine and Seiberg for SQCD \cite{ADS} 
is still missing.\footnote{Much more is known for gauge theories
containing massive adjoint and fundamental fields. See, e.g.,
\cite{EFGR,CSW,DJ} and references therein.} 

Sometime ago, Cs\'aki and Murayama showed that, for a special choice
of the numbers of colors and flavors, an effective superpotential for
the confined degrees of freedom of the Kutasov model
can be obtained by analyzing the classical constraints 
for the gauge invariant operators \cite{CM}. It is interesting that
under certain conditions the full nonperturbative superpotential is 
determined by the requirement to reproduce the 
classical constraints via its equations of motion.

In this article, we use the method developed by 
Cs\'aki and Murayama \cite{CM} to determine the low-energy effective 
nonperturbative superpotential for the model discussed by
Kutasov \cite{Ku}. Our result is a generalization of the effective 
superpotential discovered by Affleck, Dine and Seiberg 
to supersymmetric gauge theories containing matter fields in the
adjoint and fundamental representations. 
We use the results of \cite{CM} for an $\cN=1$
$SU(N_c)$ gauge theory with $N_f$ massless fundamental flavors 
and one massless adjoint field $\Phi$ with $\Wtree=h\,\tr\Phi^3$.
For $2N_f=N_c+1$, the theory confines without chiral symmetry breaking.
The effective superpotential in terms of confined degrees of
freedom was determined in \cite{CM}. Starting from this result,
we obtain the effective superpotential for $2N_f<N_c$:
\be  \label{Weff_result}
\Weff = \left(h^{N_c}\Lambda^{2(2N_c-N_f)}\over
               (\det M_2)^2\right)^{1\over N_c-2N_f}
               \Tr\left(M_1 M_2^{\ -1}\right),
\ee
where $M_1$, $M_2$ are the gauge invariant meson operators.
This generalizes the Affleck-Dine-Seiberg superpotential for SQCD.
The fact that the effective superpotential (\ref{Weff_result})
contains only one term, in an expansion in powers of 
$\Lambda^{(2N_c-N_f)/(N_c-2N_f)}$, is a consequence of the
global symmetries and physical boundary conditions in the
weak and strong coupling limits.
The situation is more complicated if the tree-level superpotential
is of the form $\Wtree=h\,\tr\Phi^{k+1}$ for $k>2$. In these
cases the effective superpotential contains several terms.
Interestingly, the additional terms do not modify the classical
constraints in the confining phase.

Using the electric-magnetic duality of the Kutasov model \cite{Ku,KS,KSS},
one can find the (leading term of) the effective superpotential
of a related theory which has the same gauge group and matter content 
but a different tree-level superpotential, which gives mass to the 
fundamental flavors and couples them to the adjoint field. 
Thus, one obtains nonperturbative information about an $\cN=1$ 
gauge theory with massless adjoint matter and massive fundamental
fields. The idea is that the effective superpotential for the electric
theory can be translated into a corresponding effective superpotential
of the dual magnetic theory. On the other hand, the magnetic theory
can be viewed as the electric theory of a different model with a 
different tree-level superpotential.

In section 2, we briefly review how the form of effective superpotentials
can be determined by symmetry considerations. We derive the dependence
on the masses and coupling parameters of the effective superpotential 
of an $\cN=1$ $SU(N_c)$ gauge theory, with massive fundamental fields 
and a massive or massless adjoint field. 
In section 3, we discuss the case where both, the adjoint and 
fundamental fields, are massless. Building upon results of Cs\'aki and
Murayama \cite{CM}, we are able to determine the exact low-energy
effective superpotential if the tree-level superpotential for the
adjoint field is cubic.
In section 4, we apply electric-magnetic Kutasov duality to the theory
with massless adjoint and fundamental fields to obtain the leading
term of the effective superpotential of the theory with massless adjoint
and massive fundamental fields.
In an appendix, we explain how to generalize some of the results of
section 3 and 4 to the case with tree-level superpotential 
$\Wtree=h\,\tr\Phi^{k+1}$ for $k>2$. We also determine the constraints
on the quantum moduli space and show that they coincide with the
classical constraints.

\section{Effective superpotentials from symmetries}
\label{Wfromsymmetry}
We start by using symmetries to constrain the form of the effective 
superpotential for an $\cN=1$ supersymmetric $SU(N_c)$ gauge theory 
with one chiral superfield $\Phi$ in the adjoint representation and 
$N_f$ quark flavors---i.e., $N_f$ chiral superfields $Q$ in the 
fundamental representation and $N_f$ chiral superfields $\bar Q$ 
in the antifundamental representation---and tree-level superpotential
\be  \label{Wtree}
\Wtree=\sum_{l=1}^{k}\Tr(m_l\,\bar Q \Phi^{l-1} Q)
       + \half\,m_\Phi \tr\Phi^2 + {\textstyle \frac{1}{k+1}}\,h 
         \tr\Phi^{k+1},
\ee
where $k$ is some integer $<N_c$, $h$ is a coupling parameter of 
mass dimension $(2-k)$ and the $m_l$ are matrix valued coupling 
parameters of mass dimension $(2-l)$.
We denoted the flavor trace by `Tr' to distinguish it
from the color trace denoted by `tr'. 
This model is a deformation of the model analyzed by Kutasov and
Schwimmer \cite{KS} and has previously been discussed in \cite{KSS}.
Although many of our results are valid for general $k$, we will
mostly be interested in the $k=2$ case, which corresponds to 
a cubic superpotential.
The form of the effective superpotential can be determined
using a method introduced by Seiberg \cite{Sei}, where one treats the
parameters as background fields that carry charges under the global
symmetries of the theory without superpotential. If the tree-level
superpotential vanishes, the theory has a large global symmetry.
The full global symmetry can be preserved by all interactions in
(\ref{Wtree}) if appropriate charges are assigned to the parameters.
As a consequence, any effective superpotential which is a function of
the parameters and gauge invariant combinations of the massless fields
has to be invariant under the full global symmetry. Some of the classical
symmetries are anomalous at the quantum level. But the effective
superpotential can be rendered invariant even under these symmetries
if one assigns appropriate charges to the dynamically generated scale 
$\Lambda$.\footnote{For a pedagogical review of these techniques see
\cite{IS2}.}

In our case, the theory with $\Wtree=0$ has a global symmetry
\be  \label{glob_sym}
G=SU(N_f)_L\times SU(N_f)_R\times U(1)_R\times U(1)_\Phi\times
                                  U(1)_A\times U(1)_B.
\ee
Under this symmetry, the fields transform as shown in table
\ref{charges_massive}.
We chose some convenient values for the Abelian charges of $\Phi$
and $Q$. The remaining charges are determined by the anomaly
freedom of $U(1)_R$ and $U(1)_B$.
For general R-charge $R_\Phi$ of the adjoint field, one finds that
the $U(1)_R$ symmetry is anomaly free if $Q$, $\bar Q$ have charge
$R_Q=1-R_\Phi{N_c\over N_f}$. The baryon symmetry $U(1)_B$ is anomaly 
free if $Q$ and $\bar Q$ have opposite charges. The anomalies of
$U(1)_\Phi$ and $U(1)_A$ can be cured by assigning charges $2N_c$
and $2N_f$ to $\Lambda^b$, where $b=2N_c-N_f$ is the one-loop 
coefficient of the beta-function.
The charges of the parameters of the non-vanishing superpotential 
(\ref{Wtree}) are then easily determined. The results are
summarized in table \ref{charges_massive}.

\begin{table}
$$\ba{c|cccccc}
          &SU(N_f)_L &SU(N_f)_R &U(1)_R &U(1)_\Phi &U(1)_A &U(1)_B\\ 
\hline
Q         &\Yfun &\one   &1-{N_c\over N_f} &0         &1      &1\\
\bar Q  &\one  &\Yfunb &1-{N_c\over N_f} &0         &1      &-1\\
\Phi      &\one  &\one   &1                &1         &0      &0\\
m_\Phi    &\one  &\one   &0                &-2        &0      &0\\
h         &\one  &\one   &-(k-1)           &-(k+1)    &0      &0\\
m_l       &\Yfunb&\Yfun  &{2N_c\over N_f}-(l-1) &-(l-1) &-2   &0\\
\Lambda^b &\one  &\one   &0                &2N_c      &2N_f   &0
\ea$$
\caption{\label{charges_massive}Charges of fields and parameters in
         (\ref{Wtree}) under the global symmetries.}
\end{table}

The effective superpotential obtained after integrating out the massive
fields $\Phi$, $Q$ and $\bar Q$ only depends on the parameters 
$m_\Phi$, $h$, $m_l$, $\Lambda$, and can be written as
\be  \label{Weff_ansatz}
\Weff \sim \Lambda^{b\,x/N_c}\prod_{l=1}^k m_l^{y_l}\, m_\Phi^w\, h^z,
\ee
for some numbers $x,y_l,w,z$, where we suppressed the flavor indices
on $m_l$ which should be contracted appropriately. Invariance under 
$U(1)_R\times U(1)_\Phi\times U(1)_A$ implies the following conditions:
\be  \label{xyz_conditions}
\sum_{l=1}^k y_l=x\,{N_f\over N_c},\quad
\sum_{l=1}^k y_l(l-1)=2x-2w-z(k+1),\quad
z+w=1.
\ee
This determines three of the $k+3$ exponents in (\ref{Weff_ansatz}).

To be more specific, let us now consider the case $k=2$ in more detail. 
One finds that the effective superpotential has to be of the form
\be \label{Weff_ktwo}
\Weff = m_\Phi\,\sum_{x,z}c_{x,z}
              \left(\Lambda^b\det m_1\right)^{x\over N_c}
        \left[\Tr\left(\left(m_2m_1^{-1}\right)^{2(x-1)-z}\right)
              +\hbox{multi-trace}\right]
              \left({h\over m_\Phi}\right)^z
\ee
to be invariant under the global symmetries. Multi-trace contributions
like $(\Tr(m_2m_1^{-1}))^n$ are allowed by the symmetries.
From the matrix model approach to determining the effective superpotential,
one knows \cite{BR} that only diagrams with at most one boundary\footnote{In
the 't Hooft double-line formalism a boundary corresponds to a trace over
fields in the fundamental representation.} contribute to the
Veneziano-Yankielowicz-Dijkgraaf-Vafa superpotential $W(S)$
\cite{VY,DV} containing the glueball superfield $S$. 
Thus each term in $W(S)$ can have at most one trace over 
flavor indices. However, integrating out $S$ from
$W(S)$ yields various multi-trace contributions in $\Weff$.
The coefficients $c_{x,z}$ are undetermined by the symmetries.
The dynamical scale $\Lambda_L$ of the low-energy theory is 
related to the high-energy scale $\Lambda$ by
\be  \label{scale_matching}
\Lambda_L^{3N_c}=\Lambda^{2N_c-N_f}\,m_\Phi^{N_c}\,\det m_1.
\ee
Inserting this relation into (\ref{Weff_ktwo}), we find
\be \label{Wefflow_ktwo}
\Weff = \Lambda_L^3\,\sum_{x,z}c_{x,z}
             \left[\Tr\left(\left(\alpha\Lambda_L^3\right)^{x-1}
                       \beta^z\right)+\hbox{multi-trace}\right],
\ee
with $\alpha=m_2m_1^{-1}m_2m_1^{-1}/m_\Phi$ and
$\beta=m_1m_2^{-1}h/m_\Phi$. The coefficients in this expansion
can be determined either by performing a matrix model computation
similar to one discussed in \cite{ACFH} or by solving the 
factorization equations of the corresponding Seiberg-Witten curve
\cite{CIV,CV,CSW}. This will be done in a separate publication \cite{KS2}.

If the adjoint field $\Phi$ is massless, we have have to set $w=0$ in
(\ref{Weff_ansatz}), which yields $z=1$ in (\ref{xyz_conditions}).
The low-energy scale $\Lambda_L$ is now given by
\be  \label{scale_matchQ}
\Lambda_L^{2N_c}=\Lambda^{2N_c-N_f}\,\det m_1.
\ee
Inserting this and the condition $z=1$ into (\ref{Weff_ktwo}), we find
\be  \label{Weff_phimassless}
\Weff = h\,\Lambda_L^3
        \sum_x c_x\left[
      \Tr\left(\left(\Lambda_L m_2m_1^{-1}\right)^{2x-3}\right)
                  +\hbox{multi-trace}\right].
\ee
Note that the effective superpotential has to be linear in $h$ to be
consistent with the global symmetries.\footnote{This is a manifestation
of the {\em linearity principle} discussed in \cite{int_in}.}
The coefficients $c_x$ cannot be determined by symmetry considerations. 
However, using the electric-magnetic duality discovered by Kutasov 
\cite{Ku}, one can show that the leading term in the expansion
(\ref{Weff_phimassless}) is $\Lambda^4$; the coefficient
$c_1$ vanishes. This will be discussed in section \ref{Wfromduality}.

\section{Effective superpotentials from classical constraints}
\label{Wfromconstraints}
We would now like to find the effective superpotential for the model
first discussed by Kutasov and Schwimmer \cite{Ku,KS}. 
This is an $\cN=1$ supersymmetric $SU(N_c)$ gauge 
theory with one adjoint field $\Phi$, $N_f$ quark flavors $Q$, $\bar Q$
and tree-level superpotential
\be  \label{WKut}
\Wtree={\textstyle \frac{1}{k+1}}\,h \tr\Phi^{k+1},
\ee
where $k$ is some integer $<N_c$ and $h$ is a parameter of mass dimension
$2-k$. The gauge invariant operators that generate the chiral ring are
\cite{KS,KSS}
\be  \label{def_MlTl}
M_l\equiv \bar Q\Phi^{l-1}Q,\quad l=1,\ldots,k,\qquad 
u_n\equiv{1\over n}\,\tr\Phi^n,\quad n=2,\ldots,k.
\ee
The charges of the elementary and composite fields are shown in 
table \ref{charges_massless}.
\begin{table}
$$\ba{c|cccccc}
          &SU(N_f)_L &SU(N_f)_R &U(1)_R &U(1)_\Phi &U(1)_A &U(1)_B\\ 
\hline
Q         &\Yfun &\one   &1-{2N_c\over(k+1)N_f} &0    &1    &{1\over N_c}\\
\bar Q  &\one  &\Yfunb &1-{2N_c\over(k+1)N_f} &0    &1    &-{1\over N_c}\\
\Phi      &\one  &\one   &{2\over k+1}     &1         &0    &0\\
u_n       &\one  &\one   &{2n\over k+1}    &n         &0    &0\\
h         &\one  &\one   &0                &-(k+1)    &0    &0\\
M_l       &\Yfun &\Yfunb &\left(-{2N_c\over N_f}+k+l\right){2\over k+1} 
                                                            &l-1 &2  &0\\
\Lambda^b &\one  &\one   &0                &2N_c      &2N_f   &0
\ea$$
\caption{\label{charges_massless}Charges of elementary and composite
         fields of the Kutasov model (\ref{WKut}).
         For convenience, we have chosen the R-charge of $\Phi$ to be
         such that $h$ is neutral under $U(1)_R$.}
\end{table}
If $kN_f\ge N_c$, then there are also baryonic operators. To define them,
we first introduce dressed quark operators:
\be  \label{def_dressed}
Q_l\equiv \Phi^{l-1}Q,\qquad \bar Q_l\equiv \bar Q\Phi^{l-1}.
\ee
The baryons carry $kN_f-N_c$ flavor indices and are given by
\bea  \label{def_baryons}
&&B^{(n_1,\ldots,n_k)} \equiv (Q_1)^{n_1}(Q_2)^{n_2}\cdots(Q_k)^{n_k},\qquad
\bar B^{(\bar n_1,\ldots,\bar n_k)} \equiv (\bar Q_1)^{\bar n_1}
               (\bar Q_2)^{\bar n_2}\cdots(\bar Q_k)^{\bar n_k},\\
&&{\rm with}\ \sum_{l=1}^k n_k=\sum_{l=1}^k \bar n_k=N_c.\nonumber
\eea
We suppressed all color and flavor indices. The integers $\{n_l\}$
label the different types of baryons. Each of them carries $kN_f-N_c$
flavor indices which we have not displayed. The exponents denote the powers
to which the dressed quarks appear. The color indices are contracted with
a rank $N_c$ epsilon-tensor. The $N_c$ flavor indices are contracted with
$k$ rank $N_f$ epsilon-tensors, leaving $kN_f-N_c$ free indices. 
(The first rank $N_f$ epsilon-tensor is contracted with $(Q_1)^{n_1}$, 
the second with $(Q_2)^{n_2}$, etc.)

Murayama and Cs\'aki observed \cite{CM} that for $kN_f=N_c+1$, the
low-energy effective theory of this model is described by confined
degrees of freedom and that the point of unbroken chiral symmetry
is not removed from the quantum moduli space. This is a generalization 
of a similar result for SQCD with $N_f=N_c+1$ by Seiberg \cite{Sei2}.
The confined degrees of freedom in the latter case obey the constraints
\be  \label{SQCD_constr}
M_{\bar\imath}^{\ j}B_j=0,\quad 
\bar B^{\bar\imath}M_{\bar\imath}^{\ j}=0,\quad
(\cof M)_j^{\ \bar\imath}=B_jB^{\bar\imath},
\ee
where $i,j$ are $SU(N_f)_L$ indices, $\bar\imath,\bar\jmath$ are
$SU(N_f)_R$ indices and the cofactor is defined by 
$\cof M=(\partial/\partial M)\det M$.
The authors of \cite{CM} developed a method how to determine the
analogous classical constraints in the Kutasov model explicitly. 
The trick is to realize that all the constraints involving quark
fields can be obtained by treating the dressed quarks (\ref{def_dressed})
as the only independent degrees of freedom. By assembling the dressed
quarks $Q_l$ in one vector of an enlarged flavor space
\be  \label{def_largeQ}
\cQ=(Q_1,\ldots,Q_k),\qquad \bar\cQ=(\bar Q_1,\ldots,\bar Q_k)
\ee
and at the same time forgetting about the $\Phi$ degrees of freedom,
one has effectively mapped the Kutasov model to SQCD with $kN_f$ flavors.
Similarly, the $kN_f$ baryons\footnote{In the case $kN_f=N_c+1$, there
are $k$ different choices for the $(n_1,\ldots,n_k)$ and each baryon
carries one $SU(N_f)$ index.} (\ref{def_baryons}) can be assembled in 
one vector of the enlarged flavor space:
\be  \label{def_largeB}
\cB=(B_1,\ldots,B_k),\qquad \bar\cB=(\bar B_1,\ldots,\bar B_k),
\ee
where $B_1\equiv B^{(N_f-1,N_f,\ldots,N_f)}$,
$B_2\equiv B^{(N_f,N_f-1,\ldots,N_f)}$, $\ldots$,
$B_k\equiv B^{(N_f,\ldots,N_f,N_f-1)}$.
Finally, one forms the enlarged meson matrix
\be  \label{def_largeM}
\cM=\bar\cQ\cQ.
\ee
Now, the constraints for the Kutasov model with $kN_f=N_c+1$ follow
directly from (\ref{SQCD_constr}) and read
\be \label{Kut_constr}
\cM_{\bar I}^{\ J}\cB_J=0,\quad 
\bar\cB^{\bar I}\cM_{\bar I}^{\ J}=0,\quad
(\cof\cM)_J^{\ \bar I}=B_JB^{\bar I},
\ee
where the capital indices run from 1 to $kN_f$. There are some additional
constraints following from the $\Phi$ equation of motion, which cannot be 
obtained in this picture of SQCD with $kN_f$ flavors. Using 
$\partial\Wtree/\partial\Phi=0$, one can show \cite{CM} that only $B_k$
and $\bar B_k$ are non-zero in (\ref{def_largeB}) and that 
$\bar Q_l Q_{l'}=0$ if $l+l'>k+1$. Thus, $\cM$, $\cB$, $\bar\cB$
are of the form
\bea  \label{form_of_MBB}
&&\cM=\left(\ba{ccccc} M_1 &M_2     &\cdots  &M_{k-1} &M_k\\
                       M_2 &\ddots  &M_{k-1} &M_k     &\\
                       \vdots  &M_{k-1} &M_k\\
                       M_{k-1} &M_k  &  &\hbox{\LARGE 0}\\
                       M_k     &        &
            \ea\right), \\
&&\cB=(0,\ldots,0,B_k),\qquad \bar\cB=(0,\ldots,0,\bar B_k). \nonumber
\eea

In the case of SQCD with $N_f=N_c+1$ flavors, it is easy to see \cite{Sei2}
that the equations of motion of the confining superpotential
\be  \label{Wconf_SQCD}
\Wconf={\bar BMB-\det M\over\Lambda^{3N_c-N_f}}
\ee
are just the classical constraints (\ref{SQCD_constr}). It has been shown
that the classical constraints do not receive any quantum corrections and
that (\ref{Wconf_SQCD}) is the exact effective superpotential at low 
energies.

Let us determine the effective superpotential for the confined degrees
of freedom of the Kutasov model \cite{CM}. Comparing the classical 
constraints (\ref{Kut_constr}) to those of SQCD, one would guess that 
the former are reproduced by a superpotential of the form 
\be  \label{Wconf_naive}
\Wconf\sim\bar\cB\cM\cB-\det\cM.
\ee
However, there are two problems with this approach. 
First, this superpotential does not have the correct R-charge. 
From the charges summarized in table \ref{charges_massless} and 
the definitions of $\cM$, $\cB$, $\bar\cB$, one finds that 
$\bar\cB\cM\cB$ has R-charge $2+{2(k-1)\over k+1}$.\footnote{One
might think that it is possible for $\bar\cB\cM\cB$ to have R-charge
2 if one chooses a different R-charge for $\Phi$ in table 
\ref{charges_massless}. However, $R_\Phi\neq2/(k+1)$ implies
$R_h\neq0$, which leads to a non-invariant superpotential since
the power of $h$ in $\Wconf$ is already fixed by $U(1)_\Phi$ invariance.}
Second, one cannot derive a superpotential for the physical 
degrees of freedom $M_l$, $B_k$, $\bar B_k$ from 
(\ref{Wconf_naive}), since inserting the constraints 
(\ref{form_of_MBB}) into (\ref{Wconf_naive}) gives
$\Wconf\sim(\det M_k)^k$. The constraints (\ref{Kut_constr}) can
only be obtained from (\ref{Wconf_naive}) by first deriving the
equations of motion and then inserting (\ref{form_of_MBB}).

These difficulties can be overcome by introducing ad-hoc a new
meson matrix
\be  \label{def_Mhat}
\widehat\cM=\left(\ba{ccccc}   &  &  &  &M_1\\
                       &\hbox{\LARGE 0}  &  &M_1  &M_2\\
                       &  &M_1  &M_2  &\vdots\\
                       &M_1  &M_2  &\ddots  &M_{k-1}\\
                       M_1  &M_2  &\cdots  &M_{k-1} &M_k
            \ea\right).
\ee
A superpotential of the form
\be  \label{Wconf_improved}
\Wconf\sim\bar\cB\widehat\cM\cB-\widehat\cM\cof\cM
\ee
has the correct R-charge, reduces to a superpotential of the physical
degrees of freedom $M_l$, $B_k$, $\bar B_k$ when the constraints
(\ref{form_of_MBB}) are inserted and reproduces the classical 
constraints (\ref{Kut_constr}) through its equations of motion
if (\ref{form_of_MBB}) is inserted into (\ref{Wconf_improved})
{\em before} the equations of motion are derived. Actually, the 
dependence of the superpotential on the fields is already fixed by
symmetry considerations as explained in the previous section. One
finds that the confining superpotential should be of the form \cite{conf}
$W\sim(\bar Q Q)^{kN_f}\Phi^{(k-1)kN_f-k+1}$
to be consistent with the global symmetries. This agrees with the form
(\ref{Wconf_improved}).

The second term in (\ref{Wconf_improved}) can be transformed using
$\widehat\cM\cof\cM\equiv\Tr(\widehat\cM^\top\!\cof\cM)
 =\det\cM\Tr(\widehat\cM\cM^{-1})$. Adding the appropriate powers
of $\Lambda$ and $h$, one finds
\be  \label{Wconf_Kut}
\Wconf=-\sigma_k\,
       {\bar\cB\widehat\cM\cB-\det\cM\Tr\left(\widehat\cM\cM^{-1}\right)
        \over h^{N_c}\Lambda^{k(2N_c-N_f)}},\quad{\rm where}\ 
        \sigma_k=(-1)^{\half k(k-1)N_f}.
\ee
This superpotential is invariant under the full global symmetry of 
table \ref{charges_massless}.
The sign factor $\sigma_k$ was added for later convenience.

For the special case $k=2$, it easy to show that
\be  \label{TrMM_ktwo} 
\Tr\left(\widehat\cM\cM^{-1}\right)=\Tr\left(M_1M_2^{\ -1}\right),
\quad \det\cM=(-1)^{N_f}(\det M_2)^2 \qquad\qquad({\rm if}\ k=2).
\ee
Inserting these identities into (\ref{Wconf_Kut}) yields the
confining superpotential in terms of $M_1$, $M_2$, $B_2$,
$\bar B_2$:
\be \label{Wconf_ktwo_explicit}
\Wconf={(-1)^{N_f+1}\bar B_2 M_2 B_2+(\det M_2)^2
        \Tr\left(M_1M_2^{\ -1}\right)
        \over h^{N_c}\Lambda^{2(2N_c-N_f)}},
\ee 
The superpotential (\ref{Wconf_ktwo_explicit}) agrees with the 
result of \cite{CM} up to the sign in front of $\bar B_k M_k B_k$. 
In the appendix, we determine the confining superpotential for
general $k$.

Note that for each fixed value of the parameters $N_f$, $N_c$, $k$,
the superpotential (\ref{Wconf_Kut}), (\ref{Wconf_ktwo_explicit}) 
contains only one definite power of $\Lambda$. One might wonder 
whether the full confining superpotential is a power series in 
$1/\Lambda^{2N_c-N_f}$. The global symmetries of table 
\ref{charges_massless} constrain possible additional terms 
to be of the form
\bea  \label{more_terms}
&&{(\bar Q Q)^{k'N_f}\Phi^{k'(k-1)N_f+(k+1-2k')}\over
  h^{k'N_f-1}\Lambda^{k'(2N_c-N_f)}} \nonumber\\
&\sim &{(\det M_k)^{k'}\left[\Tr(M_{k-k'+1}M_{k'}^{-1})
     +\cdots+\Tr(M_{2(k-k')+1}M_{k}^{-1})\right]\over
 h^{k'N_f-1}\Lambda^{k'(2N_c-N_f)}},
\eea
for some positive integer $k'$. For the expression in terms of 
the meson operators $M_l$ we assumed $k'<k$, which is satisfied
\cite{CM}
as we show below. The last term in the square bracket is only 
present if $2k'>k$. Note that only this last term is smooth
in the limit of vanishing field expectation values. Since the
't Hooft anomaly matchings for $Q$, $\bar Q$, $\Phi$ in the UV 
and $M_l$, $B_k$, $\bar B_k$ in the IR are satisfied at the
origin of the moduli space \cite{CM}, it is very unlikely that
there are additional massless degrees of freedom at the origin
of the moduli space. Thus the confining superpotential in terms
of the meson and baryon operators should be smooth at vanishing
field expectation values. This implies that only for $2k'>k$ can
there be any additional terms in the confining superpotential.%
\footnote{We thank C.~Cs\'aki and H.~Murayama for a useful
discussion on this point.}
On the other hand, the equations of motion of the confining 
superpotential should reproduce the classical constraints 
(\ref{Kut_constr}) in the limit where the fields have very 
large expectation values, $\vev{M_l}\gg\Lambda^{1+l}$. 
Thus the additional terms (\ref{more_terms}) can only arise 
for $k'\le k$ \cite{CM}. Summarizing, we find that the integer
$k'$ in (\ref{more_terms}) must satisfy
\be  \label{kprime_condition}
k<2k'\le 2k.
\ee
For $k=2$, this implies that $k'=k$ and thus the effective 
superpotential (\ref{Wconf_ktwo_explicit}) is exact \cite{CM}.
However, additional terms in the effective superpotential
are possible if $k>2$.
For $k=3$, one finds that $(\det M_3)^2/h^{2N_f-1}\Lambda^{2(2N_c-N_f)}$ 
is of the form (\ref{more_terms}) with $k'=2$ and thus consistent 
with all global symmetries. 
It would be interesting to see whether such a term is indeed present
in the confining superpotential or whether it is excluded for a 
different reason.

In \cite{CM}, it was shown that the confining superpotential
(\ref{Wconf_ktwo_explicit}) can be understood via the electric-magnetic
duality of the Kutasov model \cite{Ku,KS,KSS}. The dual magnetic
theory has gauge group $SU(\tilde N_c)$, with $\tilde N_c=kN_f-N_c$.
This means that for $kN_f=N_c+1$, the magnetic gauge group is
completely broken. The first term in (\ref{Wconf_ktwo_explicit}) 
corresponds to a tree-level term of the magnetic superpotential.
The second term in (\ref{Wconf_ktwo_explicit}) is generated by a
$k$-instanton effect in the completely higgsed magnetic gauge theory. 

The importance of the confining superpotential for the theory with 
$kN_f=N_c+1$ is that the effective superpotentials for $kN_f<N_c$
can be obtained from it by successively integrating out quark flavors.
For the case of SQCD (corresponding to $k=1$), the techniques are
reviewed in \cite{IS2}. For $k=2$, the authors of \cite{CM} computed
the effective superpotential of the theory with $2N_f=N_c-1$ by giving
mass to one quark flavor and integrating out the heavy degrees of
freedom. Generalizing their results by integrating out further
quark flavors, we find that for general $N_f$ and $N_c$, the
$k=2$ effective superpotential is given by
\bea  \label{Weff_ktwo_Kut}
\Weff &= &\sigma_k^{1\over(N_c-2N_f)}
               \left(h^{N_c}\Lambda^{2(2N_c-N_f)}\over
               \det\cM\right)^{1\over N_c-2N_f}
            \Tr\left(\widehat\cM\cM^{-1}\right)  \nonumber\\
      &= &\left(h^{N_c}\Lambda^{2(2N_c-N_f)}\over
                         (\det M_2)^2\right)^{1\over N_c-2N_f}
                         \Tr\left(M_1 M_2^{\ -1}\right).
          \qquad\qquad(k=2)
\eea
This superpotential could only be derived for values of $N_f$ and $N_c$ 
such that $N_c+1-2N_f$ is even. However, one can verify that $\Weff$ 
of (\ref{Weff_ktwo_Kut}) is invariant under the full global symmetry 
of table \ref{charges_massless}. Therefore, we conjecture that it is 
the exact effective superpotential for all $2N_f<N_c$. This generalizes 
the effective superpotential for $SQCD$ with $N_f<N_c$
found by Affleck, Dine and Seiberg \cite{ADS} to the Kutasov model. 
For $2N_f=N_c-1$ the superpotential (\ref{Weff_ktwo_Kut}) has the form
of a $2$-instanton term. 

For $2N_f<N_c-1$, one might speculate whether the effective 
superpotential is generated by gluino condensation in the unbroken 
$SU(N_c-2N_f)$ gauge theory at a generic point of the moduli space. 
This would be very similar to the situation of SQCD with $N_f<N_c-1$.
In SQCD, gluino condensation in the unbroken $SU(N_c-N_f)$ gauge
theory generates a superpotential $W=\Lambda^3_L$, which via 
the scale matching relation 
$\Lambda^{3N_c-N_f}=\Lambda^{3(N_c-N_f)}_L\det M$,
leads to the Affleck-Dine-Seiberg superpotential
$W=(\Lambda^{3N_c-N_f}/\det M)^{1/(N_c-N_f)}$.
If the same mechanism is at work in the $k=2$ Kutasov model,
we need a much more complicated scale matching relation.
If the high-energy $SU(N_c)$ gauge theory with dynamical scale 
$\Lambda$ is broken to an $SU(N_c-2N_f)$ gauge theory with scale
$\Lambda_L$ by generic expectation values for $M_1$ and $M_2$,
then the two scales should be related by
\be  \label{Kut_scalematch}
\Lambda^{2(2N_c-N_f)}=\Lambda^{3(N_c-2N_f)}_Lh^{-N_c}(\det M_2)^2
                      \left(\Tr(M_1M_2^{-1})\right)^{2N_f-N_c}
\ee
if we want that the effective superpotential (\ref{Weff_ktwo_Kut})
is generated by the gluino condensation term $W=\Lambda^3_L$.
Assuming that the gauge couplings $g$ of the $SU(N_c)$ theory and
$g_L$ of the $SU(N_c-2N_f)$ theory are matched by the relation%
\footnote{The factor 2 in the matching of the gauge couplings
$g$ and $g_L$ is related to the fact that the effective
superpotential is generated by a 2-instanton effect in the
case $2N_f=N_c-1$.}
$1/g^2(v)=1/(2g_L^2(v))$, where $v$ is the symmetry breaking scale,
one has \cite{CM2}
\be  \label{Lambda_v}
\left({\Lambda^{2N_c-N_f}\over v^{2N_c-N_f}}\right)^2
= {\Lambda^{3(N_c-2N_f)}_L\over v^{3(N_c-2N_f)}}.
\ee
This implies that the symmetry breaking scale $v$ is related to the
meson expectation values via
\be  \label{Ml_v}
\left(\det\left({M_2\over hv^3}\right)\right)^2 
= \left(hv\,\Tr(M_1M_2^{-1})\right)^{N_c-2N_f}
\ee
While this does not seem unreasonable, we do not have any 
independent argument to justify this relation.

Let us briefly consider the cases $2N_f\ge N_c$.
The authors of \cite{MP} have shown that for $2N_f=N_c$, the $k=2$ 
Kutasov model has a quantum modified moduli space with vanishing effective
superpotential. For $2N_f>N_c$, the superpotential (\ref{Weff_ktwo_Kut}) 
is still consistent with all symmetries. But it does not take into account
the baryon operators.
For $2N_f=N_c+1$, it is just the second term in the confining
superpotential (\ref{Wconf_ktwo_explicit}). Assuming that 
(\ref{Wconf_ktwo_explicit}) is still the correct superpotential for
$2N_f>N_c$ in the limit where the vacuum expectation values of 
all baryon operators vanish, we can obtain some non-trivial
information about the theories (\ref{Wtree}) with $m_\Phi=0$
by using electric-magnetic duality. This will be shown in the next
section.

\section{Effective superpotentials from duality}
\label{Wfromduality}
The $SU(N_c)$ gauge theory with one adjoint field $\Phi$, 
$N_f$ quark flavors $Q$, $\bar Q$ and tree-level superpotential
$\Wtree={h\over k+1}\tr\Phi^{k+1}$ discussed in section 
\ref{Wfromconstraints} has a dual description in terms
of magnetic variables \cite{Ku,KS,KSS}. The dual magnetic theory
has gauge group $SU(\tilde N_c)$, with $\tilde N_c=kN_f-N_c$,
one adjoint field $\tilde\Phi$, $N_f$ quark flavors $q$, $\bar q$, 
$kN_f^{\ 2}$ gauge singlets $\tilde M_l$ with charges as shown in table 
\ref{charges_mag}
\begin{table}
$$\ba{c|cccccc}
          &SU(N_f)_L &SU(N_f)_R &U(1)_R &U(1)_\Phi &U(1)_A &U(1)_B\\ 
\hline
q         &\Yfunb &\one  &1-{2\tilde N_c\over(k+1)N_f} &0  &-1 
                                                       &{1\over\tilde N_c}\\
\bar q  &\one  &\Yfun  &1-{2\tilde N_c\over(k+1)N_f} &0  &-1 
                                                       &-{1\over\tilde N_c}\\
\tilde\Phi      &\one  &\one   &{2\over k+1}     &1        &0 &0\\
\tilde h        &\one  &\one   &0                &-(k+1)   &0 &0\\
\tilde M_l  &\Yfun &\Yfunb &\left(-{2N_c\over N_f}+k+l\right){2\over k+1} 
                                                 &2(l-k) &2    &0\\
\mu         &\one   &\one  &0  &-1  &0  &0\\
\tilde\Lambda^b &\one  &\one   &0          &2\tilde N_c    &2N_f &0
\ea$$
\caption{\label{charges_mag}Charges of the fields and parameters of the 
         dual magnetic theory (\ref{Wtree_mag}).}
\end{table}
and tree-level superpotential
\be  \label{Wtree_mag}
W_{\rm mag}={\textstyle{1\over k+1}\tilde h}\tr\tilde\Phi^{k+1}\ +\ 
                     \sum_{l=1}^k \mu^{l-k} 
                     \tilde M_l\,\bar q \tilde\Phi^{k-l}q.
\ee
The scale $\mu$ had to be introduced to obtain the correct mass dimensions.
It will be related to the dynamical scales $\Lambda$ of the electric
theory and $\tilde\Lambda$ of the magnetic theory via the duality map.
Note that the gauge singlets $\tilde M_l$ have the canonical mass dimension
of a chiral superfield although they correspond to the generalized mesons 
$M_l$ of the electric theory, which have mass dimensions $1+l$.
The precise mapping is given below.

The generalized magnetic mesons $\bar q\tilde\Phi^{l-1}q$ are not
moduli of the magnetic theory; their expectation values do not correspond
to flat directions but rather are fixed by the superpotential
(\ref{Wtree_mag}). However, there are magnetic moduli corresponding
to vacuum expectation values of baryon operators.
Let us introduce dressed quark operators as we did in the electric
theory in eq.\ (\ref{def_dressed}):
\be  \label{mag_dressed}
q_l\equiv \tilde\Phi^{l-1}q,\qquad \bar q_l\equiv \bar q\tilde\Phi^{l-1}.
\ee
Now, the magnetic baryons are given by
\bea  \label{mag_baryons}
&&b^{(m_1,\ldots,m_k)} \equiv (q_1)^{m_1}(q_2)^{m_2}\cdots(q_k)^{m_k},\qquad
\bar b^{(\bar m_1,\ldots,\bar m_k)} \equiv (\bar q_1)^{\bar m_1}
               (\bar q_2)^{\bar m_2}\cdots(\bar q_k)^{\bar m_k},\\
&&{\rm with}\ \sum_{l=1}^k m_k=\sum_{l=1}^k \bar m_k=\tilde N_c.\nonumber
\eea
The integers $\{m_l\}$ label the different types of baryons. 
Each of them carries $kN_f-\tilde N_c$ flavor indices which 
we have not displayed. The exponents denote the powers to which 
the dressed quarks appear. The color indices are contracted with
a rank $\tilde N_c$ epsilon-tensor. The $\tilde N_c$ flavor indices 
are contracted with $k$ rank $N_f$ epsilon-tensors, leaving 
$kN_f-\tilde N_c$ free indices.

The precise mapping between the electric and the magnetic theory is
\cite{KSS}
\bea  \label{Kut_duality_map}
\tilde N_c &= &kN_f-N_c,\nonumber\\
\Lambda^{2N_c-N_f}\tilde\Lambda^{2\tilde N_c-N_f} &= &h^{-2N_f}\mu^{2N_f},
         \nonumber\\
\tilde h &= &-h, \nonumber\\
\tilde M_l &= &h\,\mu^{k-l-2}\,M_l, \\
b^{(m_1,\ldots,m_k)} &= &\left((-1)^{\hf k(k-1)N_f-N_c}
                           \tilde h^{kN_f}\mu^{-2\tilde N_c}
                           \Lambda^{k(2N_c-N_f)}\right)^{-1/2}
                          B^{(n_1,\ldots,n_k)},
         \nonumber\\
                &&{\rm with}\ m_l=N_f-n_{k+1-l},\nonumber\\
\tr\tilde\Phi^l &= &-\tr\Phi^l,\quad l=2,\ldots,k-1,\qquad
\tr\tilde\Phi^k\ =\ {\tilde N_c\over N_c}\tr\Phi^k.\nonumber
\eea
This mapping respects all the global symmetries of tables 
\ref{charges_massless} and \ref{charges_mag}. Again we suppressed
the flavor indices of the baryons. In the right hand side of
the fifth line of (\ref{Kut_duality_map}), $B^{(n_1,\ldots,n_k)}$
is contracted with $k$ epsilon-tensors of ranks $N_f-n_l+m_{k-l+1}$,
$l=1,\ldots,k$. 
Let us verify that both sides of the baryon mapping have the same 
mass dimension. From the definitions of $B^{(n_1,\ldots,n_k)}$ and 
$b^{(m_1,\ldots,m_k)}$ in eqs.\ (\ref{def_baryons}), 
(\ref{mag_baryons}), one finds
\bea   \label{dim_baryons}
\dim\left(B^{(n_1,\ldots,n_k)}\right) &= &\sum_{l=1}^k ln_l, \nonumber\\
\dim\left(b^{(m_1,\ldots,m_k)}\right) &= &\sum_{l=1}^k lm_l
\ =\ \sum_{l=1}^k (lN_f -(k+1-l)n_l)\nonumber\\
&= &\half(k+1)(\tilde N_c-N_c)+\dim\left(B^{(n_1,\ldots,n_k)}\right),\\
\dim\left(\tilde h^{kN_f}\mu^{-2\tilde N_c}\Lambda^{k(2N_c-N_f)}\right)
&= &kN_f(2-k)-2(kN_f-N_c)+k(2N_c-N_f) \nonumber\\
&= &(k+1)(N_c-\tilde N_c).\nonumber
\eea
Similarly, one can show that $(\ldots)^{-1/2}B^{(n_1,\ldots,n_k)}$ 
and $b^{(m_1,\ldots,m_k)}$ in eq.\ (\ref{Kut_duality_map}) have 
the same charges under all global symmetries. 
It is also instructive to verify that under two successive duality 
transformations the electric baryons are mapped onto themselves. 
Performing the duality mapping (\ref{Kut_duality_map}) twice, 
one finds 
$B^{(n_1,\ldots,n_k)}\to(-1)^{-\hf k(k-1)N_f}B^{(n_1,\ldots,n_k)}$.
The sign ambiguity is related to the possibility of field
redefinitions using the global symmetry $U(1)_B$, as explained
in \cite{KSS}.

The goal in this section is to use electric-magnetic duality in the
Kutasov model to find some nonperturbative effective superpotential
which is difficult to obtain by other methods. Using the duality
mapping (\ref{Kut_duality_map}), we can translate the effective 
superpotential (\ref{Weff_ktwo_Kut}) of the electric theory to the
magnetic theory. Written in terms of magnetic variables, the
electric superpotential (\ref{Weff_ktwo_Kut}) reads
\be  \label{Weff_ktwo_Kut_mag}
\Weff=\mu^{-1}\left(h^{\tilde N_c}\tilde\Lambda^{2(2\tilde N_c-N_f)}
                     (\det\tilde M_2)^2\right)^{1\over\tilde N_c}
              \Tr\left(\tilde M_1\tilde M_2^{\ -1}\right).
\ee
Such an effective superpotential should be generated by nonperturbative
effects in the magnetic gauge theory.
The authors of \cite{CM} showed that if the magnetic gauge group is broken 
down to $\tilde N_c=1$ by the Higgs effect, then the superpotential
(\ref{Weff_ktwo_Kut_mag}) is indeed generated by a $2$-instanton effect.
Interestingly, the only term of the magnetic tree-level superpotential
(\ref{Wtree_mag}) that remains in the completely higgsed theory with
$\tilde N_c=1$, namely $\tilde M_2\bar qq$, maps precisely to the
term proportional to $B_2 M_2 B_2$ in (\ref{Wconf_ktwo_explicit}).
From (\ref{mag_baryons}), we see that the magnetic baryons 
$b^{(1,0)}$, $\bar b^{(1,0)}$ are just given by the quark 
singlets $q$, $\bar q$ that remain light after the Higgs effect. 
Then, using (\ref{Kut_duality_map}), we find
\be  \label{Mqq_BMB_ktwo}
\tilde M_2\bar q q={(-1)^{N_f+1}\bar B_2 M_2 B_2
                            \over h^{N_c}\Lambda^{2(2N_c-N_f)}},
\ee
which is exactly the first term in (\ref{Wconf_ktwo_explicit}).
Summarizing, the first term in the confining superpotential 
(\ref{Wconf_ktwo_explicit}) of the electric theory corresponds to
the tree-level term $\tilde M_2\bar qq$ in the magnetic theory
and the second term in (\ref{Wconf_ktwo_explicit}) corresponds to
the 2-instanton term (\ref{Weff_ktwo_Kut_mag}) in the magnetic
theory (with $\tilde N_c=1$).

In general (for $\tilde N_c>1$), the effective superpotential 
(\ref{Weff_ktwo_Kut_mag})
also contains contributions including baryon operators. We have
not determined these contributions. However, (\ref{Weff_ktwo_Kut_mag})
is correct in the limit where the vacuum expectation values of all
baryon operators vanish.

Comparing the magnetic singlets $\tilde M_l$ in (\ref{Wtree_mag}) with
the coupling parameters $m_l$ in (\ref{Wtree}), we see that for
$m_l=\mu^{1-l}\tilde M_{k+1-l}$ and $m_\Phi=0$ the two superpotentials agree
if we replace $q$, $\bar q$, $\tilde\Phi$ by $Q$, $\bar Q$, $\Phi$ 
(not using the duality map but just renaming the elementary fields). 
Thus, (\ref{Weff_ktwo_Kut_mag}) gives the effective superpotential for
the case of general $m_l$ in (\ref{Wtree}) but massless adjoint field.%
\footnote{We implicitly assumed that the $m_l$ are background fields.
The low-energy result for constant $m_l$ is obtained by taking these 
background fields to be very heavy and effectively replacing them by 
their expectation values.}
Explicitly, one finds that the effective superpotential for the theory
(\ref{Wtree}) with $m_\Phi=0$ and $k=2$ is given by
\bea  \label{Weffdualresult_ktwo}
\Weff &= &\left(h^{N_c}\Lambda^{2(2N_c-N_f)}
                     (\det m_1)^2\right)^{1\over N_c}
          \Tr\left(m_2 m_1^{\ -1}\right)
%          \ +\ \cO\left(\Lambda^{3(2N_c-N_f)}\right)
\nonumber\\
      &= &h\,\Lambda_L^4\,
             \Tr\left(m_2 m_1^{\ -1}\right),
%          \ +\ \cO\left(\Lambda^6_L\right),
\eea
where we dropped the tildes over $N_c$ and $\Lambda$ and
we used the scale matching condition
$\Lambda_L^{2N_c}=\Lambda^{2N_c-N_f}\det m_1$.
When $m_1$, $m_2$ are considered to be constants, the right
hand side of (\ref{Weffdualresult_ktwo}) represents only the 
leading term in an expansion in $\Lambda_L$.
Higher powers of $\Lambda_L$ are generated upon integrating
out the quantum fluctuations of the background fields $m_l$.
Also, the electric-magnetic duality used to arrive at the 
result (\ref{Weffdualresult_ktwo}) is only valid close to 
the IR fixed point, which again implies that it is only
the leading term.
The important result of this analysis is that the leading term 
in the effective superpotential (\ref{Weffdualresult_ktwo}) is
proportional to $\Lambda^4_L$. The 1-instanton term proportional
to $\Lambda^2_L$, which was expected by naive dimensional analysis
is not present. 
This result can be verified by factorizing the corresponding 
Seiberg-Witten curve of this model \cite{KS2}.

\section{Conclusions and discussion}
In this paper, we derived the effective superpotential for an 
$\cN=1$ $SU(N_c)$ gauge theory with one massless adjoint matter 
field and $N_f$ massless fundamental flavors and a cubic superpotential 
for the adjoint field. This generalizes the well-known effective
superpotential for SQCD with $N_f<N_c$ found by Affleck, Dine and
Seiberg to the model first discussed by Kutasov. 
The global symmetries alone are not enough to fix the precise form of 
the superpotential of the Kutasov model. However the additional 
requirement that the equations of motion reproduce the classical 
constraints on the moduli space in the special case $2N_f=N_c+1$
uniquely determines the effective superpotential.
If the tree-level superpotential for the adjoint field is quartic
or of higher order, the global symmetries and the constraints on
the moduli space are not sufficient to determine the effective
superpotential. However, one can obtain the leading term in an
expansion in powers of $\Lambda$, as we have shown in the appendix.
Moreover, we were able to determine the exact constraints on the
quantum moduli space for $kN_f=N_c+1$ and we showed that they
coincide with the classical constraints.

Using Kutasov duality, the effective superpotential for the theory
with massless adjoint and fundamental fields and cubic superpotential
can be written in terms of magnetic variables. The magnetic theory
has almost the same matter content but the rank of the gauge
group is different and there are additional gauge singlets, which
are elementary fields. These gauge singlets couple to the quarks
and to the adjoint field via Yukawa-like couplings and effectively 
provide masses for the quarks.
Thus, via Kutasov duality, we found the leading term of the effective
superpotential of a supersymmetric gauge theory with one massless 
adjoint field and $N_f$ massive quark flavors and a superpotential 
which couples the quarks to the adjoint field.

Recently, Dijkgraaf and Vafa discovered a surprising link between
effective superpotentials of supersymmetric gauge theories and
the effective potential of an associated bosonic matrix model 
\cite{DV}.
This is very interesting since it allows one to obtain 
nonperturbative results by doing a perturbative calculation.
It is a powerful method to compute the effective superpotential
of $\cN=1$ gauge theories with massive matter in tensor and 
fundamental representations. However, it is difficult to use
this method if the gauge theory contains massless fields.\footnote{See 
however \cite{Feng}, where the case of massless {\it fundamental} 
fields is discussed.}
More recently, it is understood why such a simplification can arise as 
a consequence of (super)symmetries by considering Ward identities associated 
to the Konishi anomaly \cite{CDSW} (see also \cite{DVG}), 
which turns out to be the same as the loop equation in the matrix model.  
The loop equation is identical to the minimization of the superpotential, 
which in turn requires that all periods of the generating 1-form  $T$ on 
the Riemann surface defined by the superpotential are  integers. 
By Abel's theorem, the 1-form must be a derivative of a meromorphic 
function $\psi$. Finally, the condition that $\psi$ be single valued 
on the reduced Riemann surface is the factorization of the Seiberg-Witten 
curve of the original $\cN=2$ theory to the reduced curve defined by 
the superpotential. This completes the solution to the problem and 
corresponds to  extending the earlier results of Vafa and his 
collaborators \cite{CIV,CV}. 

If we can show this effective superpotential agrees with the result from 
an explicit factorization of the corresponding Seiberg-Witten curve, one 
can  provide further evidence for the validity of the electric-magnetic 
duality discovered by Kutasov. This program is under progress and will 
be published in separate paper \cite{KS2}.

\vskip20mm
\centerline{\bf Acknowledgements}

We would like to thank C.~Cs\'aki, H.~Murayama and R.~Tatar for helpful 
discussions.
The work of MK is funded by the Deutsche Forschungsgemeinschaft. 
The work of SJS is supported by by KOSEF Grant 1999-2-112-003-5.

\vskip1cm

\begin{appendix}
\section{Effective superpotentials for $k>2$}
In this appendix, we would like to generalize some of the formulae
that were only derived for the $k=2$ Kutasov model to the situation
where $\Wtree={h\over k+1}\,\tr\Phi^{k+1}$, for general $k$.

Generalizing, eq.\ (\ref{TrMM_ktwo}), one finds
\bea  \label{Tr_MM}
\Tr\left(\widehat\cM\cM^{-1}\right) &= &(k-1)\Tr\left(M_1M_k^{\ -1}\right)
\ -\ (k-2)\Tr\left(M_2M_k^{\ -1}M_{k-1}M_k^{\ -1}\right) \nonumber\\
&&\hspace*{-1cm}
  +(k-3)\left[\Tr\left(M_3M_k^{\ -1}M_{k-1}M_k^{\ -1}M_{k-1}M_k^{\ -1}\right)
       -\Tr\left(M_3M_k^{\ -1}M_{k-2}M_k^{\ -1}\right)\right]\nonumber\\
&&\hspace*{-1cm}
 -(k-4)\Big[\Tr\left(M_4M_k^{\ -1}M_{k-1}M_k^{\ -1}M_{k-1}M_k^{\ -1}
                           M_{k-1}M_k^{\ -1}\right) \nonumber\\
&&\hspace*{-1cm}
 \qquad\qquad-\Tr\left(M_4M_k^{\ -1}M_{k-2}M_k^{\ -1}M_{k-1}M_k^{\ -1}\right)
\nonumber\\
&&\hspace*{-1cm}
 \qquad\qquad-\Tr\left(M_4M_k^{\ -1}M_{k-1}M_k^{\ -1}M_{k-2}M_k^{\ -1}\right)
\nonumber\\
&&\hspace*{-1cm}
 \qquad\qquad+\Tr\left(M_4M_k^{\ -1}M_{k-3}M_k^{\ -1}\right)\Big]
\nonumber\\
&&\hspace*{-1cm}
 +(k-5)\Big[\cdots\Big]\quad +-\ \cdots \quad +(-1)^k\Big[\cdots\Big].
\eea
The determinant of the meson matrix  in (\ref{form_of_MBB}) is now
given by
\be  \label{det_cM}
\det\cM=(-1)^{\half k(k-1)N_f}(\det M_k)^k.
\ee
Inserting these results into (\ref{Wconf_Kut}), we find
\be \label{Wconf_explicit}
\Wconf={-(-1)^{\half k(k-1)N_f}\bar B_k M_k B_k+(\det M_k)^k
        \left[(k-1)\Tr\left(M_1M_k^{\ -1}\right)\ +\ \ldots\ \right]
        \over h^{N_c}\Lambda^{k(2N_c-N_f)}},
\ee 
where the `$\ldots$' represent the terms proportional to $(k-2)$,
$(k-3)$, $\ldots$, in (\ref{Tr_MM}). The superpotential 
(\ref{Wconf_explicit}) agrees with the result of \cite{CM} up to
the sign in front of $\bar B_k M_k B_k$. The method described above
provides us with a simple algorithm to compute the explicit
coefficients of the various terms for the cases with $k>2$.

Successively integrating out quark flavors from (\ref{Wconf_explicit}), 
we obtain an effective superpotential for general $k$, $N_f$ and $N_c$:
\bea  \label{Weff_Kut}
\Weff &= &(-1)^{k(k-1)N_f\over 2(N_c-kN_f)}
               \left(h^{N_c}\Lambda^{k(2N_c-N_f)}\over
               \det\cM\right)^{1\over N_c-kN_f}
            \Tr\left(\widehat\cM\cM^{-1}\right)  \nonumber\\
      &= &\left(h^{N_c}\Lambda^{k(2N_c-N_f)}\over
                         (\det M_k)^k\right)^{1\over N_c-kN_f}
                \Big[(k-1)\Tr\left(M_1 M_k^{\ -1}\right)\ +\ \ldots\ \Big].
\eea

However, as explained in section \ref{Wfromconstraints}, for $k>2$,
the confining superpotential (\ref{Wconf_explicit}) contains
additional terms of the form (\ref{more_terms}). Thus, the effective
superpotential (\ref{Weff_Kut}) is not the full answer but only
contains the terms with the highest power in $\Lambda$.

As a consequence of the additional terms (\ref{more_terms}) in the
confining superpotential (\ref{Wconf_explicit}), the classical
constraints (\ref{Kut_constr}) might be modified quantum mechanically.
This would be in sharp contrast to the case of SQCD and to the $k=2$
Kutasov model, where the classical constraints remain exact in
the full quantum theory. However, a more detailed examination shows 
that the classical constraints do not receive any quantum corrections.
To see this, let us first consider the confining superpotential for 
$k=3$ and $3N_f=N_c+1$. It consists of of a term proportional to
$\Lambda^{-3(2N_c-N_f)}$ as given in (\ref{Wconf_explicit}) and
one additional term proportional to $\Lambda^{-2(2N_c-N_f)}$ as
given in (\ref{more_terms}), with $k'=2$.
\bea  \label{Wconf_kthree}
\Wconf &= &{(-1)^{N_f+1}\bar B_3 M_3 B_3+(\det M_3)^3
        \left[2\Tr\left(M_1M_3^{\ -1}\right)\ -\ 
                 \Tr\left(M_2M_3^{\ -1}M_2M_3^{\ -1}\right)\right]
        \over h^{N_c}\Lambda^{3(2N_c-N_f)}} \nonumber\\
       &&+\ a\,{(\det M_3)^2\over h^{2N_f-1}\Lambda^{2(2N_c-N_f)}},
\eea
where $a$ is some unknown relative coefficient. The constraints on
the quantum moduli space are obtained from the equations of motion
of (\ref{Wconf_kthree}). From $\partial\Wconf/\partial M_1=0$, one
finds that $\det M_3=0$ on the quantum moduli space. Thus, the
quantum constraints are given by
\be  \label{constraints_kthree}
\bar B_3M_3 = 0 = M_3 B_3,\qquad
(-1)^{N_f+1}\bar B_3 B_3 = (\cof M_3) \Tr\left(M_2(\cof M_3)
                                           M_2(\cof M_3)\right).
\ee
This coincides with the classical constraints. The situation in the
case of general $k$ is very similar. The additional terms 
(\ref{more_terms}) do not contain a term proportional to $M_1$. And
from (\ref{Tr_MM}), one finds that the only term that contains $M_1$
in (\ref{Wconf_explicit}) is the one proportional to 
$(\det M_k)^k\,\Tr(M_1M_k^{-1})$. Thus, the equation of motion for
$M_1$ implies that $\det M_k=0$ on the quantum moduli space. This,
in turn, leads to the following quantum constraints:
\be  \label{constraints_k}
\bar B_kM_k = 0 = M_k B_k,\qquad
(-1)^{\hf k(k-1)N_f+k}\bar B_k B_k = 
(\cof M_k) \Tr\left((M_{k-1}\cof M_k)^{k-1}\right),
\ee
which again coincides with the classical constraints.

The effective superpotential (\ref{Weff_Kut}) can be written in
terms of magnetic variables using the duality mapping 
(\ref{Kut_duality_map}).
\be  \label{Weff_Kut_mag}
\Weff=\mu^{1-k}\left(h^{\tilde N_c}\tilde\Lambda^{k(2\tilde N_c-N_f)}
                     (\det\tilde M_k)^k\right)^{1\over\tilde N_c}
     \Big[(k-1)\Tr\left(\tilde M_1\tilde M_k^{\ -1}\right)\ +\ \ldots\ \Big].
\ee
Again we can replace the magnetic singlets by their expectation values
$m_l=\mu^{1-l}\tilde M_{k+1-l}$ as we did at the end of section
\ref{Wfromduality}.
If we also replace $q$, $\bar q$, $\tilde\Phi$
by $Q$, $\bar Q$, $\Phi$ (not using the duality map but
just renaming the elementary fields),
we obtain
\bea  \label{Weffdualresult}
\Weff &= &\left(h^{N_c}\Lambda^{k(2N_c-N_f)}
                     (\det m_1)^k\right)^{1\over N_c}
     \Big[(k-1)\Tr\left(m_k m_1^{\ -1}\right)\ +\ \ldots\ \Big]
\nonumber\\
      &= &h\,\Lambda_L^{2k}\,
      \Big[(k-1)\Tr\left(m_k m_1^{\ -1}\right)\ +\ \ldots\ \Big].
\eea
However this result is not as useful as the corresponding one for
the $k=2$ case since we neglected the additional terms
(\ref{more_terms}) in (\ref{Wconf_explicit}) which lead to terms
with lower powers of $\Lambda$ in (\ref{Weff_Kut}) and thus to 
additional terms with lower powers of $\Lambda_L$ in 
(\ref{Weffdualresult}).

\end{appendix}

\end{document}